\documentclass[aps,PRB,twocolumn,superscriptaddress,preprintnumbers]{revtex4-2}
\usepackage[T1]{fontenc}
\usepackage{times}
\usepackage{graphicx,color}
\usepackage{amsfonts,amsmath,amssymb,amsbsy}
\usepackage[colorlinks=true, linkcolor=blue, citecolor=blue, urlcolor=blue]{hyperref}
\usepackage{float}

\begin{document}

\title{Analytic approach to quantum metric and optical conductivity \\
in Dirac models with parabolic mass in arbitrary dimensions}
\author{Motohiko Ezawa}
\affiliation{Department of Applied Physics, The University of Tokyo, 7-3-1 Hongo, Tokyo
113-8656, Japan}

\begin{abstract}
The imaginary part of the quantum geometric tensor is the Berry curvature,
while the real part is the quantum metric. Dirac fermions derived from a
tight-binding model naturally contains a mass term $m(k)$ with parabolic
dispersion, $m(k)=$ $m+uk^{2}$. However, in the Chern insulator based on
Dirac fermions, only the sign of the mass $m$ is relevant. Recently, it was
reported that the quantum metric is observable by means of the optical
conductivity, which is significantly affected by the parabolic coefficient $%
u $. We analytically obtain the quantum metric and the optical conductivity
in the Dirac Hamiltonian in arbitrary dimensions, where the Dirac mass has
parabolic dispersion. The optical conductivity at the band-edge frequency
significantly depends on the dimensions. We also make an analytical study
on the quantum metric and the optical conductivity in the
Su-Schrieffer-Heeger model, the Qi-Wu-Zhang model and the Haldane model. The
optical conductivity is found to be quite different between the topological
and trivial phases even when the gap is taken identical.
\end{abstract}

\date{\today }
\maketitle

\section{Introduction}

There is a rapid growing interest in quantum geometry in condensed matter
physics\cite{Matsuura,Ozawa,Onishi} especially in the context of optical
conductivity\cite{Cook,Juan,Ahn,Holder,Bhalla,Ahn2,Ghosh} and electric
nonlinear conductivity\cite%
{Sodeman,MaNature,CWang,KamalDas,AGao,NWang,Kaplan,Sala,YFang}. The geometry
of quantum states in a parametrized Hilbert space is described by the
quantum geometric tensor. Namely, the distance between two quantum states in
the parameter space defines the quantum geometric tensor. The quantum metric
is its symmetric real part\cite{Berry,BerryRev,Provost,Ma}, while the Berry
curvature is its antisymmetric imaginary part. The Berry curvature leads to
topological insulators, as is well known. A typical example is the Chern
insulator\cite{Haldane} characterized by the Chern number $C$. The Chern
number is given by the integration of the Berry curvature over the whole
Brillouin zone\cite{Berry,Thouless}. On the other hand, the quantum geometry
is less explored\cite%
{Matsuura,Cook,Juan,Ahn,Ozawa,Holder,Bhalla,Ahn2,Ghosh,Onishi,Sala},
although there are some experimental observations in such as superconducting
qubit\cite{Tan}, anomalous Hall effect\cite{Gian}, qubit in diamond\cite{Yu}%
, optical active system\cite{Ren}, organic microcavity\cite{Liao}, flat-band
superconductivity\cite{Tian} and optical Raman lattice\cite{Yi}. It is
recently pointed out\cite{Ghosh} that the quantum metric is related to the
optical conductivity. The parabolic coefficient $u$\ in the Dirac mass term $%
m\left( k\right) =m+uk^{2}$\ affects the quantum metric and the optical
conductivity in the two-dimensional system\cite{Ghosh}. However, it is yet
to be explored what will happen in other dimensions. In addition, the
extension to the tight-binding model is also yet to be made.

A two-dimensional Dirac Hamiltonian provides us with a typical system to
realize a Chern insulator, where $C=\text{sgn}\left( m/2\right) $ for each
Dirac cone as in many Chern insulators. There are an even number of Dirac
cones in the tight-binding model owing to the Nielsen-Ninomiya theorem\cite%
{Nielsen}, which results in the quantized Chern number in total. The Dirac
fermions derived from a tight-binding model naturally contain a mass term $%
m(k)$ with parabolic dispersion, $m(k)=m+uk^{2}$. However, the parabolic
term $uk^{2}$ is irrelevant to the Chern number $C=\text{sgn}\left(
m/2\right) $.

In this paper, we analytically derive the quantum metric and the optical
conductivity of the Dirac Hamiltonian, where the Dirac mass term has a
parabolic dispersion in arbitrary dimensions. We have\ found that both the
quantum metric and the optical conductivity have dependences on the
parabolic coefficient $u$. The optical conductivity at the band-edge
frequency has a strong dimensional dependence. It diverges in the
one-dimensional system. It is nonzero and finite in the two-dimensional
system. It is zero in systems for more than two dimensions. We also present
analytic results of quantum metric and optical conductivity based on
tight-binding models. We explicitly investigate the Su-Schrieffer-Heeger
(SSH) model\cite{SSH}, which is the simplest model of a topological
insulator, the Qi-Wu-Zhang (QWZ) model\cite{QWZ}, which is a typical model
of the Chern insulator on the square lattice, and the Haldane model\cite%
{Haldane}, which is a typical model of the Chern insulator on the honeycomb
lattice. The optical conductivity is found to be quite different between the
topological and trivial phases even when the gap is taken identical.
Furthermore, we study the quantum metric and the optical conductivity in a
three-dimensional lattice Dirac model.

\section{Quantum metric and optical absorption}

We study the Dirac Hamiltonian in an $N$-dimensional space defined by\cite%
{FuKane,Sch,Ryu,Chiu} 
\begin{equation}
H\left( \mathbf{k}\right) =\sum_{j=0}^{N}d_{j}\left( \mathbf{k}\right)
\Gamma _{j},  \label{DiracH}
\end{equation}%
where $d_{j}\left( \mathbf{k}\right) $ is the Dirac vector and $\Gamma _{j}$
is the Gamma matrix satisfying $\left\{ \Gamma _{i},\Gamma _{j}\right\}
=2\delta _{ij}$. The Dirac mass term $m\left( \mathbf{k}\right) $ is given
by $d_{0}\left( \mathbf{k}\right) $, i.e., $m\left( \mathbf{k}\right)
=d_{0}\left( \mathbf{k}\right) $.

The quantum metric $g_{\mu \nu }$ is in general defined by the quantum
distance\cite{Provost,Souza,Matsuura,Resta},%
\begin{equation}
ds^{2}=1-\left\vert \left\langle \partial _{k_{\mu }}\psi \left( \mathbf{k}%
\right) \left\vert \partial _{k_{\nu }}\psi \left( \mathbf{k}+\delta \mathbf{%
k}\right) \right\rangle \right. \right\vert ^{2}=g_{\mu \nu }\left( \mathbf{k%
}\right) \delta k_{\mu }\delta k_{\nu },
\end{equation}%
where%
\begin{align}
g_{\mu \nu }\left( \mathbf{k}\right) =& \text{Re}[\left\langle \partial
_{k_{\mu }}\psi \left( \mathbf{k}\right) \left\vert \partial _{k_{\nu }}\psi
\left( \mathbf{k}\right) \right\rangle \right.  \notag \\
& -\left\langle \partial _{k_{\mu }}\psi \left( \mathbf{k}\right) \left\vert
\psi \left( \mathbf{k}\right) \right\rangle \right. \left\langle \psi \left( 
\mathbf{k}\right) \left\vert \partial _{k_{\nu }}\psi \left( \mathbf{k}%
\right) \right\rangle \right. ].
\end{align}%
In the Dirac model (\ref{DiracH}), it is explicitly given by\cite%
{Matsuura,Gers,Onishi,WChen2024}%
\begin{equation}
g_{\mu \nu }\left( \mathbf{k}\right) =2^{N-3}\left( \partial _{k_{\mu }}%
\mathbf{n}\right) \cdot \left( \partial _{k_{\nu }}\mathbf{n}\right) ,
\label{Gmn}
\end{equation}%
where $n_{j}\left( \mathbf{k}\right) =d_{j}\left( \mathbf{k}\right) /E\left( 
\mathbf{k}\right) \ $is the normalized Dirac vector with the energy%
\begin{equation}
E\left( \mathbf{k}\right) =\sqrt{\sum_{j=0}^{N}d_{j}^{2}\left( \mathbf{k}%
\right) }.
\end{equation}%
The diagonal component of the quantum metric is positive, 
\begin{equation}
g_{\mu \mu }\left( \mathbf{k}\right) =2^{N-3}\left( \partial _{k_{\mu }}%
\mathbf{n}\right) ^{2}.
\end{equation}%
Details on the quantum metric are summarized in Appendix A.

The quantum metric is observable in terms of the real part of the optical
conductivity\cite{WChen2022,Onishi,Souza,Sousa,Ghosh}, 
\begin{equation}
\text{Re}\left[ \sigma _{xx}\left( \omega \right) \right] =\pi e^{2}\omega
\int d\mathbf{k\;}g_{xx}\left( \mathbf{k}\right) \delta \left( \varepsilon
_{+}\left( \mathbf{k}\right) -\varepsilon _{-}\left( \mathbf{k}\right)
-\hbar \omega \right) ,  \label{Conduc}
\end{equation}%
where $\sigma _{xx}$ is the diagonal optical conductivity, $\hbar \omega $
is the photon energy, $\varepsilon _{\pm }\left( \mathbf{k}\right) $ is the
energy dispersion of the occupied ($-$) and valence ($+$) bands, and $%
g_{xx}\left( \mathbf{k}\right) $ is the quantum metric. It follows from Eq.(%
\ref{Conduc}) that the optical absorption is zero when the photon energy is
smaller than the band gap $\Delta $ ($\hbar \omega <\Delta $), where the
corresponding frequency $\omega _{0}=\Delta /\hbar $ is the band-edge
frequency. The relation between the optical conductivity and the quantum
metric is summarized in Appendix B.

\section{Dirac model with parabolic mass term}

We study the Dirac Hamiltonian (\ref{DiracH}) in $N$ dimensions with the
Dirac vector defined by%
\begin{equation}
d_{0}=m+uk^{2},\qquad d_{j}=v\eta _{j}k_{j},  \label{DiracVector}
\end{equation}%
where $m$ is the Dirac mass, $k_{j}$ is the momentum with $1\leq j\leq N$, $%
k^{2}=\sum_{j=1}^{N}k_{j}^{2}$, $u$ is the parabolic coefficient, $v$ is the
velocity and $\eta _{j}=\pm 1$ represents the helicity of the Dirac cone.
The parabolic dispersion in the Dirac mass term naturally arises from the
tight-binding model. We explicitly derive it in the QWZ model and the
Haldane model later. The two-dimensional model with $u=1$ and $\eta _{j}=1$
was studied in the previous work\cite{Ghosh}.

The energy dispersion is given by%
\begin{equation}
E=\sqrt{v^{2}k^{2}+\left( m+uk^{2}\right) ^{2}}.
\end{equation}%
The band gap is $\Delta =2\left\vert m\right\vert $, which occurs at $k=0$
for $u\geq -v^{2}/2m$. For simplicity, we only consider the case $%
u>-v^{2}/2m $.

By inserting (\ref{DiracVector}) into (\ref{Gmn}), the quantum metric $%
g_{xx}\left( \mathbf{k}\right) $ is calculated as 
\begin{equation}
g_{xx}\left( \mathbf{k}\right) =\frac{2^{N-3}v^{2}}{E^{2}}\left( 1-k_{x}^{2}%
\frac{4mu+v^{2}}{E^{2}}\right) .  \label{gxxk}
\end{equation}%
A detailed derivation is shown in Appendix C. The integration of $%
g_{xx}\left( \mathbf{k}\right) $ over the whole angle gives%
\begin{align}
g_{xx}\left( k\right) & \equiv \int g_{xx}\left( \mathbf{k}\right) \frac{J}{%
k^{N-1}}d\theta _{1}d\theta _{2}\cdots d\theta _{N-1}  \notag \\
\!\!\!\!& =\frac{2^{N-3}v^{2}N\pi ^{N/2}}{E^{2}\Gamma \left( \frac{N}{2}%
+1\right) }\left( 1-\frac{k^{2}}{N}\frac{4mu+v^{2}}{E^{2}}\right) ,
\label{gxx}
\end{align}%
where $J$ is the Jacobian shown in Appendix D, and $\Gamma $ is the gamma
function. The ($N-1$)-sphere coordinate is summarized in Appendix D. We note
that there is no dependence on $\eta _{j}$. At the Dirac point, the quantum
metric is given by%
\begin{equation}
g_{xx}\left( 0\right) =\frac{2^{N-3}v^{2}N\pi ^{N/2}}{m^{2}\Gamma \left( 
\frac{N}{2}+1\right) },  \label{gxx0}
\end{equation}%
which diverges for the massless Dirac Hamiltonian with $m=0$.

With the use of the relation 
\begin{equation}
\partial _{k}E\left( k\right) =k\frac{v^{2}+2u\left( m+uk^{2}\right) }{%
E\left( k\right) },
\end{equation}%
the optical conductivity is calculated as%
\begin{align}
\text{Re}\left[ \sigma _{xx}\left( \omega \right) \right] & =\frac{\pi
e^{2}\omega }{2}\xi _{N}\int_{0}^{\infty }k^{N-1}dk\delta \left( 2E\left(
k\right) -\hbar \omega \right) \frac{g_{xx}\left( k\right) }{\left\vert
\partial _{k}E\left( k\right) \right\vert }  \notag \\
& =\pi e^{2}\omega k_{0}^{N-2}\frac{g_{xx}\left( k_{0}\right) E\left(
k_{0}\right) }{\left\vert v^{2}+2u\left( m+uk_{0}^{2}\right) \right\vert },
\label{Sxx}
\end{align}%
where%
\begin{equation}
k_{0}=\frac{1}{\sqrt{2}u}\sqrt{-\left( 2mu+v^{2}\right) +\sqrt{v^{4}+u\left(
4mv^{2}+u\left( \hbar \omega \right) ^{2}\right) }}  \label{k0}
\end{equation}%
is the solution of 
\begin{equation}
2E\left( k_{0}\right) -\hbar \omega =0  \label{Ew}
\end{equation}%
and $\xi _{1}=2$ and $\xi _{N}=1$ for $N\geq 2$.

We observe typical behaviors in the following two cases: First, at the
band-edge frequency $\hbar \omega =2\left\vert m\right\vert $, we have $%
k_{0}=0$. In the vicinity of the band edge, Eq.(\ref{Sxx}) with the aid of
Eq.(\ref{gxx0}) yields%
\begin{eqnarray}
\text{Re}\left[ \sigma _{xx}\left( \frac{2\left\vert m\right\vert }{\hbar }%
\right) \right] &=&\pi e^{2}\xi _{N}\frac{2\left\vert m\right\vert }{\hbar }%
k_{0}^{N-2}\frac{2^{2N-5}\pi ^{N-1}v^{2}}{m^{2}\left\vert
v^{2}+2um\right\vert }  \notag \\
&\propto &k_{0}^{N-2}.
\end{eqnarray}%
It diverges in one dimension 
\begin{equation}
\lim_{k_{0}\rightarrow 0}\text{Re}\left[ \sigma _{xx}\left( \frac{%
2\left\vert m\right\vert }{\hbar }\right) \right] \propto
\lim_{k_{0}\rightarrow 0}\frac{1}{k_{0}}=\infty .
\end{equation}%
It is finite in two dimensions. It is zero for $N\geq 3$, 
\begin{equation}
\lim_{k_{0}\rightarrow 0}\text{Re}\left[ \sigma _{xx}\left( \frac{%
2\left\vert m\right\vert }{\hbar }\right) \right] \propto
\lim_{k_{0}\rightarrow 0}k_{0}^{N-2}=0.
\end{equation}%
Second, at the high frequency limit $\omega \rightarrow \infty $, the
momentum is%
\begin{equation}
\lim_{\omega \rightarrow \infty }k_{0}=\sqrt{\frac{\hbar \omega }{2u}},
\end{equation}%
by solving $\hbar \omega =2uk^{2}$. Hence, the optical conductivity at the
high frequency is given by%
\begin{align}
\lim_{\omega \rightarrow \infty }\text{Re}\left[ \sigma _{xx}\left( \omega
\right) \right] & =\frac{2^{N-3}N\pi ^{3N/2-1}v^{4}\xi _{N}}{\hbar \Gamma
\left( \frac{N}{2}+1\right) u^{2}}\left( \frac{\hbar \omega }{2}\right)
^{N/2-2}  \notag \\
& \propto \omega ^{N/2-2}.
\end{align}%
Hence, it decays as a function of $\omega $ for $N\leq 3$.

\subsection{One-dimensional model}

\begin{figure}[t]
\centerline{\includegraphics[width=0.48\textwidth]{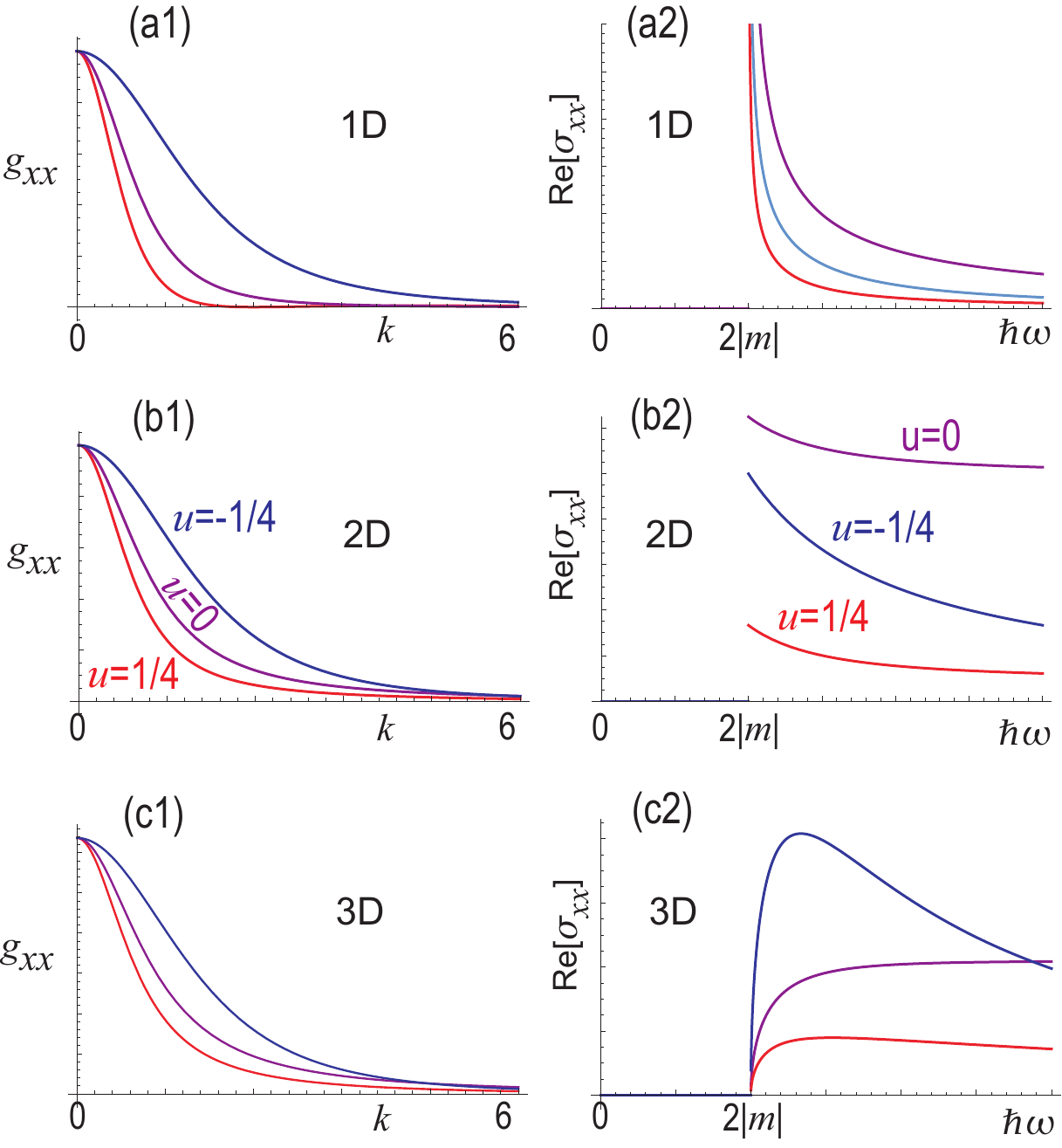}}
\caption{Dirac model. (a1), (b1) and (c1) Quantum metric as a function of $k$%
. (a2), (b2) and (c2) Optical conductivity $\text{Re}\left[ \protect\sigma %
_{xx}\right] $ as a function of $\hbar \protect\omega $. (a1) and (a2) One
dimension. (b1) and (b2) Two dimensions. (c1) and (c2) Three dimensions. Red
curves indicate $u=t/4$, purple curves indicate $u=0$, and blue curves
indicate $u=-t/4$. We have set $m=t$ and $v=t$.}
\label{FigDiracOpt}
\end{figure}

In the one-dimensional Dirac Hamiltonian, the quantum metric (\ref{gxx}) is
simply given by%
\begin{equation}
g_{xx}\left( k\right) =v^{2}\frac{\left( m-uk^{2}\right) ^{2}}{2E^{4}}.
\end{equation}%
The quantum metric\ is shown as a function of $k$ in Fig.\ref{FigDiracOpt}%
(a1). The real part of the optical conductivity (\ref{Sxx}) is obtained as%
\begin{equation}
\text{Re}\left[ \sigma _{xx}\right] =\frac{2\pi e^{2}}{\hbar \left( \hbar
\omega /2\right) ^{2}}\frac{1}{k_{0}}\frac{\left( m-uk_{0}^{2}\right) ^{2}}{%
\left\vert v^{2}+2u\left( m+uk_{0}^{2}\right) \right\vert }.
\end{equation}%
The optical conductivity is shown as a function of $\omega $ in Fig.\ref%
{FigDiracOpt}(a2). At the band-edge frequency $\hbar \omega =2\left\vert
m\right\vert $, we have $k_{0}=0$. Hence, the optical conductivity at the
band-edge frequency diverges%
\begin{equation}
\text{Re}\left[ \sigma _{xx}\left( \frac{2\left\vert m\right\vert }{\hbar }%
\right) \right] \propto \lim_{k_{0}\rightarrow 0}\frac{1}{k_{0}}=\infty ,
\end{equation}%
as shown in Fig.\ref{FigDiracOpt}(a2).

\subsection{Two-dimensional model}

In the two-dimensional Dirac Hamiltonian, the quantum metric (\ref{gxx}) is
simply given by%
\begin{equation}
g_{xx}\left( k\right) =\int_{0}^{\pi }g_{xx}\left( \mathbf{k}\right) d\theta
=\frac{\pi v^{2}}{E^{2}}\left( 1-\frac{k^{2}}{2}\frac{4mu+v^{2}}{E^{2}}%
\right) .
\end{equation}%
The quantum metric as a function of $k$\ is shown in Fig.\ref{FigDiracOpt}%
(b1). The real part of the optical conductivity is obtained as%
\begin{equation}
\text{Re}\left[ \sigma _{xx}\left( \omega \right) \right] =\frac{e^{2}\pi
^{2}v^{2}}{\hbar \left( \hbar \omega /2\right) ^{2}}\frac{\left(
vk_{0}\right) ^{2}+2\left( m^{2}+u^{2}k_{0}^{4}\right) }{\left\vert
v^{2}+2u\left( m+uk_{0}^{2}\right) \right\vert }.
\end{equation}%
The optical conductivity (\ref{Sxx}) is shown as a function of $\omega $ in
Fig.\ref{FigDiracOpt}(b2). At the band-edge frequency $\hbar \omega
=2\left\vert m\right\vert $, we have $k_{0}=0$. It is given by 
\begin{equation}
\text{Re}\left[ \sigma _{xx}\left( \frac{2\left\vert m\right\vert }{\hbar }%
\right) \right] =\frac{e^{2}\pi ^{2}v^{2}}{\hbar \left( \hbar \omega
/2\right) ^{2}}\frac{2m^{2}}{\left\vert v^{2}+2um\right\vert },
\end{equation}%
which is consistent with the previous study\cite{Onishi} in the case of $u=1$%
.

\subsection{Three-dimensional model}

In the three-dimensional Dirac Hamiltonian, the quantum metric (\ref{gxx})
is simply given by%
\begin{align}
g_{xx}\left( k\right) & =\int_{0}^{\pi }\sin \theta d\theta g_{xx}\left( 
\mathbf{k}\right) \int_{0}^{2\pi }d\phi  \notag \\
& =4\pi \frac{v^{2}}{E^{2}}\left( 1-\frac{k^{2}}{3}\frac{4mu+v^{2}}{E^{2}}%
\right) .
\end{align}%
The quantum metric as a function of $k$\ is shown in Fig.\ref{FigDiracOpt}%
(c1). The real part of the optical conductivity (\ref{Sxx}) is obtained as%
\begin{align}
\text{Re}\left[ \sigma _{xx}\left( \omega \right) \right] =& \pi e^{2}\omega
k_{0}\frac{4\pi \frac{v^{2}}{E^{2}}\left( 1-k_{0}^{2}\frac{4mu+v^{2}}{3E^{2}}%
\right) E\left( k_{0}\right) }{\left\vert v^{2}+2u\left( m+uk_{0}^{2}\right)
\right\vert }  \notag \\
=& \frac{8\pi ^{2}e^{2}v^{2}k_{0}}{3\hbar \left( \hbar \omega /2\right) ^{2}}%
\frac{2v^{2}k_{0}^{2}+3m^{2}+2muk_{0}^{2}+3u^{2}k_{0}^{4}}{\left\vert
v^{2}+2u\left( m+uk_{0}^{2}\right) \right\vert }.
\end{align}%
The optical conductivity (\ref{Sxx}) is shown as a function of $\omega $ in
Fig.\ref{FigDiracOpt}(c2). The optical conductivity is zero at the band-edge
frequency,%
\begin{equation}
\text{Re}\left[ \sigma _{xx}\left( \frac{2\left\vert m\right\vert }{\hbar }%
\right) \right] =0
\end{equation}%
for $N$ dimensions with $N\geq 3$.

\section{Dirac model without parabolic mass}

We study the Dirac Hamiltonian (\ref{FigHaldaneMetric}) with $u=0$. The
quantum metric is simply given by%
\begin{equation}
g_{xx}\left( \mathbf{k}\right) =\frac{2^{N-3}v^{2}}{E^{2}}\left( 1-k_{x}^{2}%
\frac{v^{2}}{E^{2}}\right) ,
\end{equation}%
and%
\begin{equation}
g_{xx}\left( k\right) =\frac{2^{N-3}v^{2}N\pi ^{N/2}}{E^{2}\Gamma \left( 
\frac{N}{2}+1\right) }\left( 1-\frac{v^{2}k^{2}}{NE^{2}}\right) .
\end{equation}%
They are shown by purple curves in Fig.\ref{FigDiracOpt}(a1), (b1) and (c1).

The real part of the optical conductivity is obtained as%
\begin{align}
& \text{Re}\left[ \sigma _{xx}\left( \omega \right) \right]  \notag \\
& =\frac{2^{N-2}v^{2}Ne^{2}\pi ^{N/2+1}\xi _{N}}{\hbar \Gamma \left( \frac{N%
}{2}+1\right) }\left( 1-\frac{v^{2}k_{0}^{2}}{N\left( \frac{\hbar \omega }{2}%
\right) ^{2}}\right) k_{0}^{N-2},  \label{SigmaDirac}
\end{align}%
where we used the relation%
\begin{equation}
2\partial _{k}E\left( k\right) =2k\frac{v^{2}}{E\left( k\right) }.
\end{equation}

The momentum (\ref{k0}) is singular at $u=0$. However, we may solve (\ref{Ew}%
) by setting $u=0$ to find that%
\begin{equation}
k_{0}=\frac{\sqrt{\left( \hbar \omega /2\right) ^{2}-m^{2}}}{v}.
\end{equation}%
By substituting this for $k_{0}$ in Eq.(\ref{SigmaDirac}), we obtain%
\begin{align}
\text{Re}\left[ \sigma _{xx}\left( \omega \right) \right] & =\frac{%
2^{N-2}e^{2}\pi ^{N/2+1}\xi _{N}}{\hbar \Gamma \left( \frac{N}{2}+1\right)
v^{N-4}\left( \frac{\hbar \omega }{2}\right) ^{2}}  \notag \\
& \!\!\!\!\times \left( \left( N-1\right) \left( \frac{\hbar \omega }{2}%
\right) ^{2}+m^{2}\right) \left( \left( \hbar \omega /2\right)
^{2}-m^{2}\right) ^{N/2-1}.
\end{align}%
It does not depend on the sign of $m$ in contrast to the case of $u\neq 0$
as in Eq.(\ref{Sxx}). It is shown by purple curves in Fig.\ref{FigDiracOpt}%
(a2), (b2) and (c2).

\begin{figure}[t]
\centerline{\includegraphics[width=0.48\textwidth]{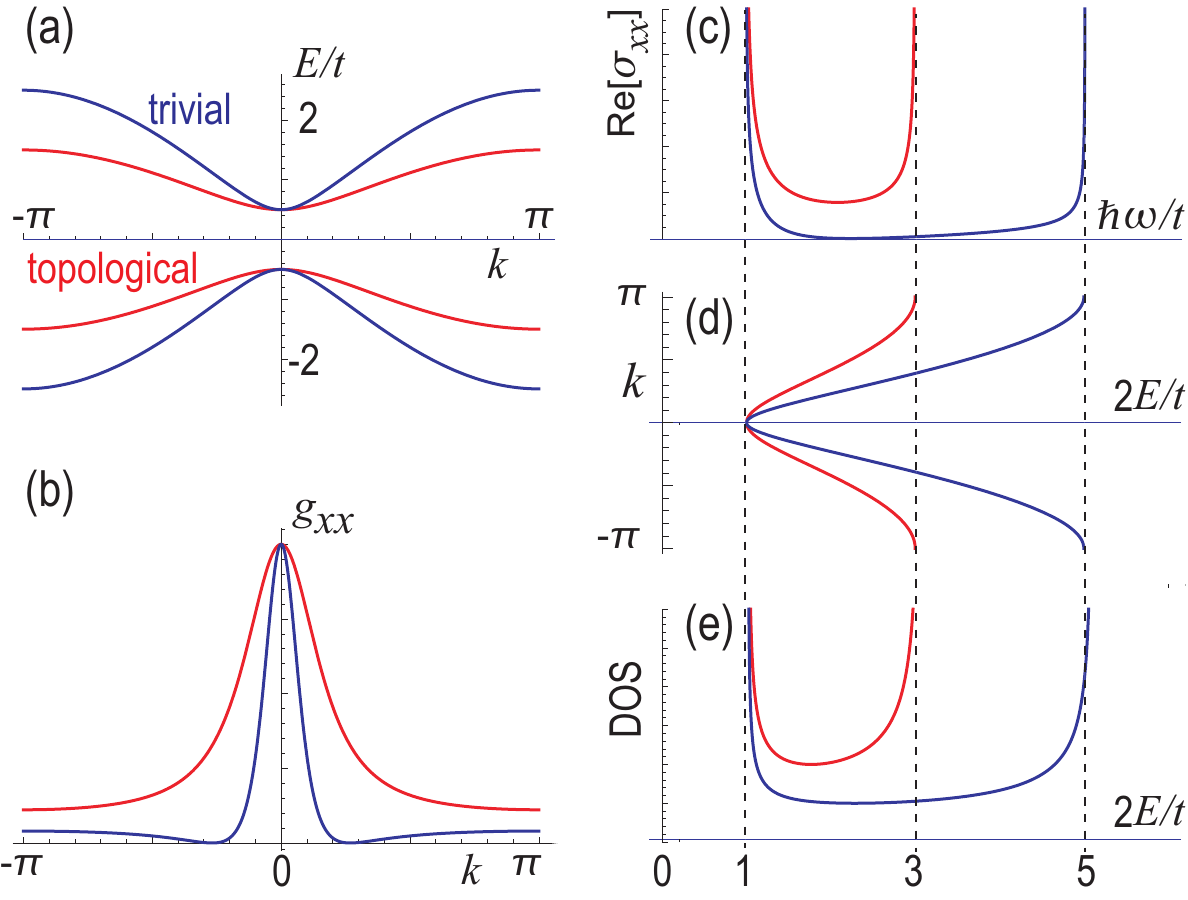}}
\caption{SSH model. (a) Energy spectrum. The horizontal axis is the momentum 
$k$. The vertical axis is the energy $E/t$. (b) Quantum metric $g_{xx}\left(
k\right) $. The horizontal axis is the momentum $k$. (c) Optical
conductivity Re$\left[ \protect\sigma _{xx}\right] $. The horizontal axis is 
$\hbar \protect\omega $. (d) The momentum $k$\ as a function of the energy $%
2E/t$. This figure is identical to the energy spectrum in (a) but for the
orientation and the scale. (e) DOS as a function of the energy $2E/t$ Red
color indicates $m_{0}=0.5t$, where the system is topological. Blue color
indicates $m_{0}=1.5t$, where the system is trivial. We have set $v=t$. }
\label{FigQSSH}
\end{figure}

\section{Tight-binding models on the hypercubic lattice}

Next, we study the $N$-dimensional tight-binding model on the hypercubic
lattice, where the Dirac vector is given by\cite{WChen2020,WChen2024}%
\begin{equation}
d_{0}=m_{0}-t\sum_{j=1}^{N}\cos k_{j},\quad d_{j}=v\sin k_{j},
\end{equation}%
where $1\leq j\leq N$ and $m_{0}$ is the model parameter. In the vicinity of
the $\Gamma $ point, we have%
\begin{equation}
m=m_{0}-Nt,\qquad u=t/2.
\end{equation}%
We explicitly discuss several models in what follows.

\subsection{Kitaev model}

We study the tight-binding model in one dimensional chain,%
\begin{equation}
H=d_{0}\sigma _{x}+d_{x}\sigma _{y},
\end{equation}%
where the Dirac vector is given by%
\begin{equation}
d_{0}=m_{0}-t\cos k,\quad d_{x}=v\sin k,  \label{Wilson}
\end{equation}%
and $\sigma _{j}$ is the Pauli matrix. A typical model is the Kitaev $p$%
-wave topological superconductor model\cite{Kitaev01}.

The quantum metric is given by%
\begin{equation}
g_{xx}\left( k\right) =\frac{v^{2}\left( t-m_{0}\cos k\right) ^{2}}{4E\left(
k\right) ^{4}}.
\end{equation}%
The optical conductivity is calculated as%
\begin{equation}
\text{Re}\left[ \sigma _{xx}\left( \omega \right) \right] =\frac{\pi
e^{2}v^{2}\left( t-m_{0}\cos k_{0}\right) ^{2}}{2\hbar \left( \frac{\hbar
\omega }{2}\right) ^{2}\left( m_{0}t+\left( v^{2}-t^{2}\right) \cos
k_{0}\right) \sin k_{0}},
\end{equation}%
where we have used%
\begin{equation}
2\partial _{k}E\left( k\right) =2\frac{\left( m_{0}t+\left(
v^{2}-t^{2}\right) \cos k\right) \sin k}{E\left( k\right) }
\end{equation}%
with%
\begin{equation}
k_{0}=\arccos \frac{2m_{0}t-\sqrt{4v^{2}\left( m_{0}^{2}-t^{2}+v^{2}\right)
+\left( t^{2}-v^{2}\right) \hbar \omega }}{2\left( t^{2}-v^{2}\right) }.
\label{k0Kita}
\end{equation}

\subsection{SSH model}

In the SSH\ model\cite{SSH}, we have $v=t$ in Eq.(\ref{Wilson}). In this
case, the momentum (\ref{k0Kita}) is singular. However, we may solve (\ref%
{Ew}) by setting $v=t$ to find that 
\begin{equation}
k_{0}=\arccos \frac{\left( m_{0}^{2}+t^{2}\right) -\left( \frac{\hbar \omega 
}{2}\right) ^{2}}{2m_{0}t}.
\end{equation}%
By substituting this for $k_{0}$ in Eq.(\ref{Sxx}), the optical conductivity
is simplified as 
\begin{equation}
\text{Re}\left[ \sigma _{xx}\left( \omega \right) \right] =\frac{\pi e^{2}}{%
4\hbar \left( \frac{\hbar \omega }{2}\right) ^{2}}\frac{\left(
t^{2}-m_{0}^{2}+\left( \frac{\hbar \omega }{2}\right) ^{2}\right) ^{2}}{%
\sqrt{\left( 2m_{0}t\right) ^{2}-\left( \left( m_{0}^{2}+t^{2}\right)
-\left( \frac{\hbar \omega }{2}\right) ^{2}\right) ^{2}}}.
\end{equation}%
We study two typical cases, $m_{0}=0.5t$ and $m_{0}=1.5t$, where the system
is topological and trivial, respectively.

We show the energy spectrum in Fig.\ref{FigQSSH}(a), where the gap is given
by $2\left\vert m_{0}-t\right\vert $. The quantum metric is shown in Fig.\ref%
{FigQSSH}(b). The optical conductivity $\text{Re}\left[ \sigma _{xx}\left(
\omega \right) \right] $ is shown in Fig.\ref{FigQSSH}(c). It diverges at $%
\hbar \omega =2\left\vert m_{0}-t\right\vert $, which is consistent with the
Dirac model as shown in Fig.\ref{FigDiracOpt}(a2). In addition, it diverges
at $2\left\vert m_{0}+t\right\vert $.

We explain the structure of the optical conductivity in Fig.\ref{FigQSSH}(c)
as follows. We show Fig.\ref{FigQSSH}(d) which is identical to the energy
spectrum Fig.\ref{FigQSSH}(a) except for the orientation and the scale. The
band-edge frequency in the optical conductivity coincides with the band gap $%
2\left\vert m_{0}-t\right\vert $\ of the energy spectrum, and the sharp peak
in the optical conductivity emerges when the the gap energy $2E$\ becomes
flat with respect to $k$\ in Fig.\ref{FigQSSH}(d). We also show the density
of states (DOS) in Fig.\ref{FigQSSH}(e), where the sharp peak in the optical
conductivity is found to be due to the van-Hove singularity.

\subsection{QWZ model}

As a typical example of the Chern insulator on square lattice, we study the
QWZ model\cite{QWZ},%
\begin{equation}
H=\left[ m_{0}-t\left( \cos k_{x}+\cos k_{y}\right) \right] \sigma
_{z}+v\left( \sigma _{x}\sin k_{x}+\sigma _{y}\sin k_{y}\right) ,
\end{equation}%
where the Dirac vector is given by 
\begin{align}
d_{0}& =m_{0}-t\left( \cos k_{x}+\cos k_{y}\right) ,  \notag \\
d_{x}& =v\sin k_{x},\quad d_{y}=v\sin k_{y}.
\end{align}%
The Dirac vector is obtained as%
\begin{equation}
d_{0}=m_{0}+\xi t+\frac{\zeta _{x}k_{x}^{2}+\zeta _{y}k_{y}^{2}}{2}t,\quad
d_{x}=v\eta _{x}k_{x},\quad d_{y}=v\eta _{y}k_{y},
\end{equation}%
where $\eta _{x}=1$, $\eta _{y}=1$, $\xi =-2$, $\zeta _{x}=1$ and $\zeta
_{y}=1$ at the $\Gamma $ point; $\eta _{x}=-1$, $\eta _{y}=-1$, $\xi =2$, $%
\zeta _{x}=-1$ and $\zeta _{y}=-1$ at the $M$ point; $\eta _{x}=1$, $\eta
_{y}=-1$, $\xi =0$, $\zeta _{x}=-1$ and $\zeta _{y}=1$ at the $X$ point; $%
\eta _{x}=-1$, $\eta _{y}=1$, $\xi =0$, $\zeta _{x}=1$ and $\zeta _{y}=-1$
at the $Y$ point.

\begin{figure}[t]
\centerline{\includegraphics[width=0.48\textwidth]{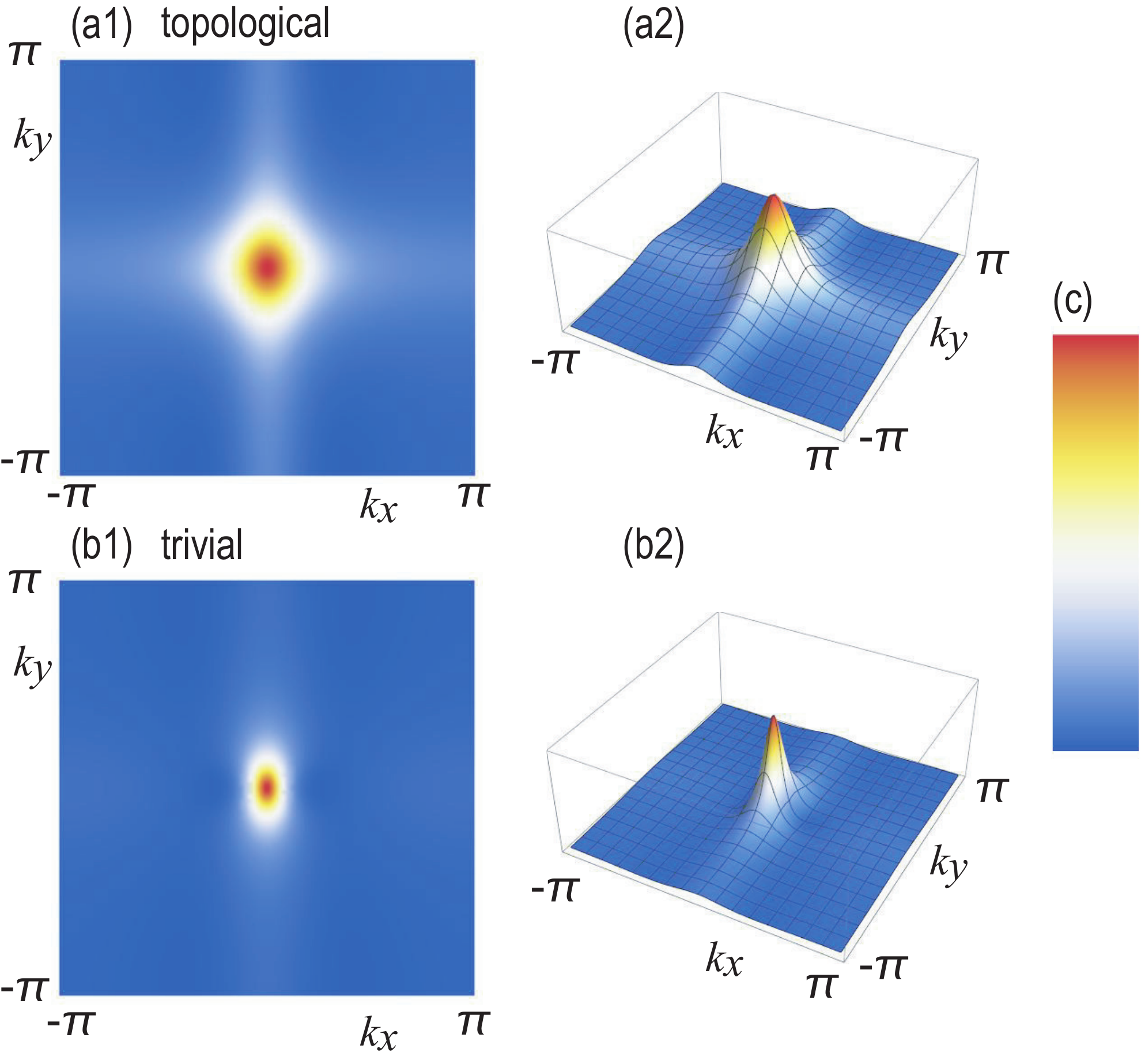}}
\caption{QWZ model. Quantum metric $g_{xx}$ in the $\left(
k_{x},k_{y}\right) $ plane. (a1) and (a2) $m_{0}=1.5t$, where the system is
topological. (b1) and (b2) $m_{0}=2.5t$, where the system is trivial. (a1)
and (b1) Density plot of $g_{xx}$. (a2) and (b2) Bird's eye's view of $%
g_{xx} $. (c) Colar palette for (a1), (a2), (b1) and (b2). }
\label{FigSqMetric}
\end{figure}

\begin{figure}[t]
\centerline{\includegraphics[width=0.48\textwidth]{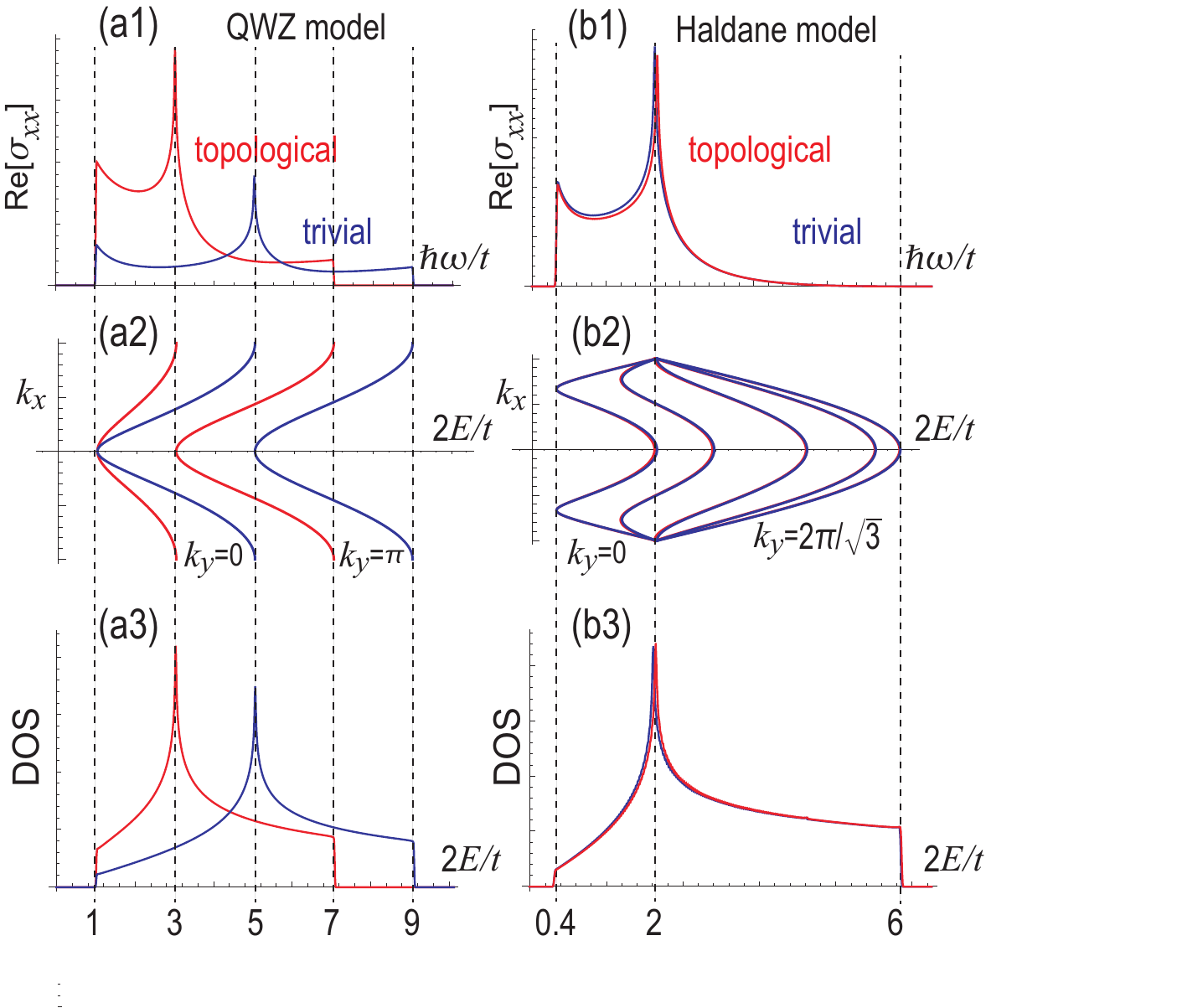}}
\caption{ (a1), (a2) and (a3) QWZ model. Red color indicates $m_{0}=1.5t$,
where the system is topological. Blue color indicates $m_{0}=2.5t$, where
the system is trivial. (b1), (b2) and (b3) Haldane model. Blue curves
indicate $m_{0}=0.2t$ and $\protect\lambda =0$, where the system is trivial.
Red curves indicate $m_{0}=0$ and $\protect\lambda =0.2t$, where the system
is topological. (a1) and (b1) Optical conductivity Re$\left[ \protect\sigma %
_{xx}\right] $. The horizontal axis is $\hbar \protect\omega $. (a2)
Momentum $k_{x}$ for $k_{y}=0$ and $\protect\pi $. (b2) Momentum $k_{x}$ for 
$k_{y}=0$ and $a\protect\pi /(2\protect\sqrt{3})$ with $a=0,1,2,3$ and $4$.
The horizontal axis is the energy $2E/t$. (a3) and (b3) DOS as a function of
the energy $2E/t$.}
\label{FigQOpt}
\end{figure}

\begin{figure}[t]
\centerline{\includegraphics[width=0.48\textwidth]{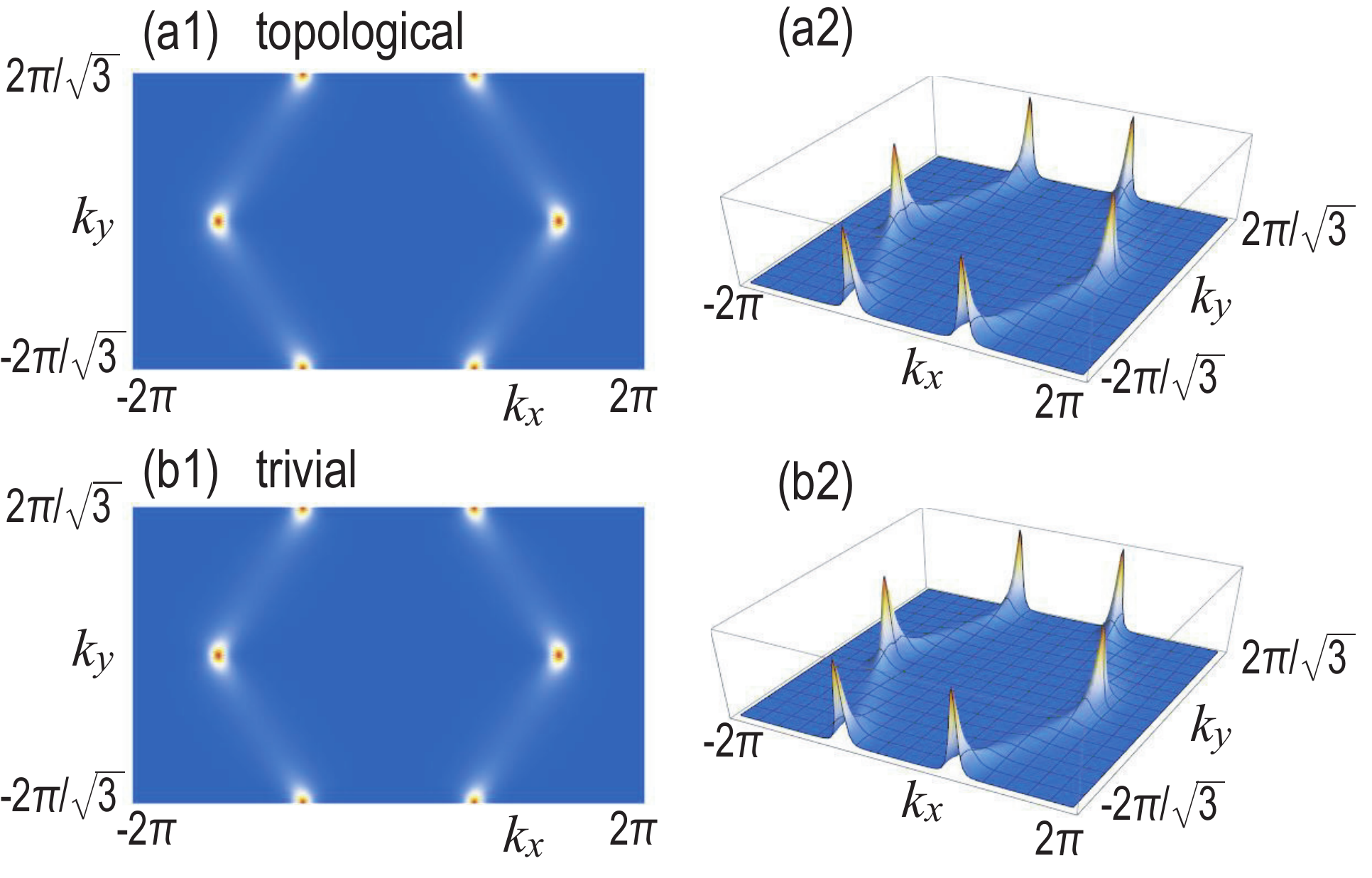}}
\caption{Haldane model. Quantum metric $g_{xx}$ in the $\left(
k_{x},k_{y}\right) $ plane. (a1) and (a2) $m_{0}=0$ and $\protect\lambda %
=0.2t$, where the system is topological. (b1) and (b2) $m_{0}=0.2t$ and $%
\protect\lambda =0$, where the system is trivial. (a1) and (b1) Density plot
of $g_{xx}$. (a2) and (b2) Bird's eye's view of $g_{xx}$. Color palette is
given by Fig.\protect\ref{FigSqMetric}(c).}
\label{FigHaldaneMetric}
\end{figure}

We consider two typical cases, $m_{0}=\left( 2-\alpha \right) t$ and $%
m_{0}=\left( 2+\alpha \right) t$ with $0<\alpha <1$, where the band gap is
present at the $\Gamma $ point with the gap $2\alpha t$, which are identical
between the two cases. The system is topological in the case of $%
m_{0}=\left( 2-\alpha \right) t$, while it is trivial in the case of $%
m_{0}=\left( 2+\alpha \right) t$. The quantum metric is shown in Fig.\ref%
{FigSqMetric} for $\alpha =0.5$. The optical conductivity is shown in Fig.%
\ref{FigQOpt}(a1). The optical conductivity is drastically different between
the two phases although the band gaps are identical. It is understood as
follows. We assume $v=t$ for simplicity.

The optical conductivity at the band-edge frequency $\hbar \omega
=2\left\vert m\right\vert $ is proportional to 
\begin{equation}
\frac{2m^{2}}{\left\vert v^{2}+2um\right\vert }=\frac{2\left( -\alpha
t\right) ^{2}}{\left\vert t^{2}-\alpha t^{2}\right\vert }=\frac{2\alpha ^{2}%
}{\left\vert 1-\alpha \right\vert }
\end{equation}%
in the case $m_{0}=\left( 2-\alpha \right) t$, and 
\begin{equation}
\frac{2m^{2}}{\left\vert v^{2}+2um\right\vert }=\frac{2\left( \alpha
t\right) ^{2}}{\left\vert t^{2}+\alpha t^{2}\right\vert }=\frac{2\alpha ^{2}%
}{\left\vert 1+\alpha \right\vert }
\end{equation}%
in the case $m_{0}=\left( 2+\alpha \right) t$. The ratio is 
\begin{equation}
\frac{1+\alpha }{1-\alpha }=3
\end{equation}%
for $\alpha =0.5$, which is significantly large.

We explain the structure of the optical conductivity in Fig.\ref{FigQOpt}%
(a1) as in the case of the SSH model. We show a figure which is identical to
the energy spectrum except for the orientation and the scale in Fig.\ref%
{FigQOpt}(a2). The band-edge frequency in the optical conductivity coincides
with the band gap $2\left\vert m_{0}-2t\right\vert $\ of the energy
spectrum, and the sharp peak in the optical conductivity emerges when the
the gap energy $2E$\ becomes flat with respect to $k_{x}$\ in Fig.\ref%
{FigQOpt}(a2). We also show the density of states (DOS) in Fig.\ref{FigQOpt}%
(a3), where the sharp peak in the optical conductivity is due to the
van-Hove singularity.

\subsection{Haldane model}

Next, we study the Haldane model on the honeycomb lattice\cite{Haldane},%
\begin{equation}
H=d_{0}\sigma _{z}+d_{x}\sigma _{x}+d_{y}\sigma _{y},
\end{equation}%
where the Dirac vector is given by 
\begin{align}
d_{0}& =m_{0}+\frac{\lambda }{3\sqrt{3}}\left( \sin k_{x}-\sum_{\pm }\sin 
\frac{k_{x}\pm \sqrt{3}k_{y}}{2}\right) ,  \notag \\
d_{x}& =t\left( \cos \frac{k_{y}}{\sqrt{3}}+2\cos \frac{2k_{y}}{\sqrt{3}}%
\cos \frac{k_{x}}{2}\right) ,  \notag \\
d_{y}& =t\left( -\sin \frac{k_{y}}{\sqrt{3}}+2\sin \frac{2k_{y}}{\sqrt{3}}%
\cos \frac{k_{x}}{2}\right) .
\end{align}

There exist Dirac cones at the $K$ point ($\eta =1$) and the $K^{\prime }$
point ($\eta =-1$), where $(k_{x},k_{y})=(4\pi \eta /3,0)$. We define the
momentum $k_{x}^{\prime }=k_{x}-4\pi \eta /3$ measured from the $K$ or $%
K^{\prime }$ point, and we replace $k$ in Eq.(\ref{DiracVector}) with $%
k^{\prime }$. The Dirac mass $m$, the velocity $v$ and parabolic coefficient 
$u$ are given by%
\begin{equation}
m=m_{0}-\eta \lambda ,\quad v=\sqrt{3}t/2,\quad u=\eta \lambda /4.
\end{equation}

We study two cases, $(m_{0},\lambda )=\left( \alpha t,0\right) $ and $%
(m_{0},\lambda )=\left( 0,\alpha t\right) $, where the gaps are identical.
The system is trivial in the case of $(m_{0},\lambda )=\left( \alpha
t,0\right) $, while it is topological in the case of $(m_{0},\lambda
)=\left( 0,\alpha t\right) $. The quantum metric is shown in Fig.\ref%
{FigHaldaneMetric} for $\alpha =0.2$. The quantum metrics are almost
identical between the two cases. The optical conductivity is shown in Fig.%
\ref{FigQOpt}(b1). The difference is tiny between the two cases. It is
understood as follows. The optical conductivity at the band-edge frequency
is 
\begin{equation}
\frac{2m^{2}}{\left\vert v^{2}+2um\right\vert }=\frac{2\left( \alpha
t\right) ^{2}}{\left\vert \left( \frac{\sqrt{3}t}{2}\right) ^{2}\right\vert }%
=\frac{2\alpha ^{2}}{3/4}
\end{equation}%
in the case $(m_{0},\lambda )=\left( \alpha t,0\right) $, while it is%
\begin{equation}
\frac{2m^{2}}{\left\vert v^{2}+2um\right\vert }=\frac{2\left( \alpha
t\right) ^{2}}{\left\vert \left( \frac{\sqrt{3}t}{2}\right) ^{2}+2\frac{%
\alpha t}{4}\alpha t\right\vert }=\frac{2\alpha ^{2}}{3/4+\alpha ^{2}/2}
\end{equation}%
in the case $(m_{0},\lambda )=\left( 0,\alpha t\right) $. The ratio is%
\begin{equation}
\frac{3/4+\alpha ^{2}/2}{3/4}=1.027
\end{equation}%
for $\alpha =0.2$, which is very tiny.

The structure of the optical conductivity in Fig.\ref{FigQOpt}(b1) is
understood as in the case of the QWZ model. Namely, the band-edge frequency
in the optical conductivity coincides with the band gap of the energy
spectrum as in Fig.\ref{FigQOpt}(b2), and the sharp peak in the optical
conductivity is due to the van-Hove singularity in the DOS in Fig.\ref%
{FigQOpt}(b3).

\subsection{Three-dimensional lattice Dirac model}

\begin{figure}[t]
\centerline{\includegraphics[width=0.48\textwidth]{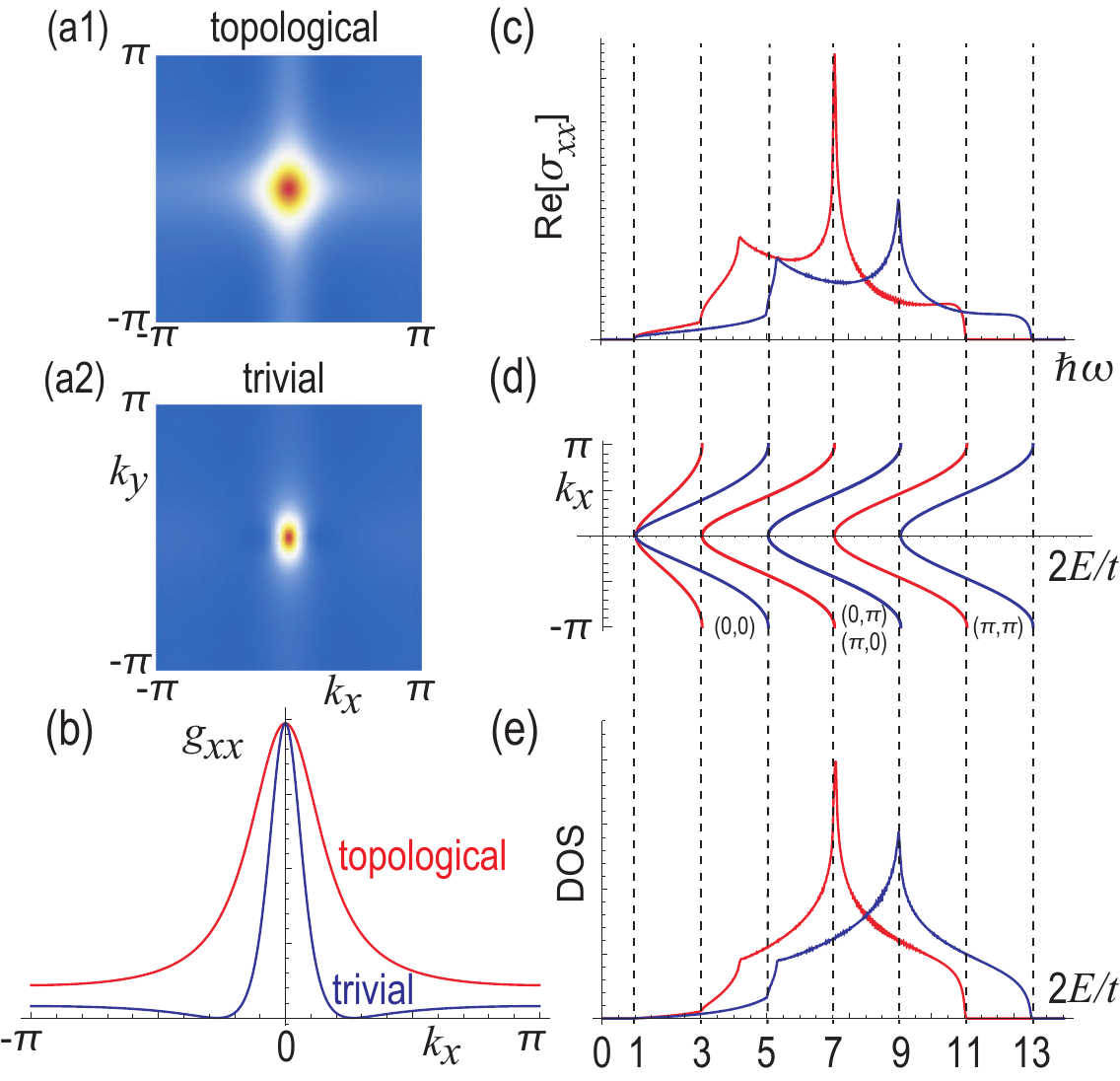}}
\caption{3D lattice Dirac model. (a1) and (a2) Quantum metric in the ($k_{x}$%
,$k_{y}$) plane along the $k_{z}=0$ plane. (a1) $m_{0}=2.5t$, where the
system is topological. (a2) $m_{0}=3.5t$, where the system is trivial. Color
palette is given by Fig.\protect\ref{FigSqMetric}(c). (b) Quantum metric
along the $k_{x}$ axis, where $k_{y}=k_{z}=0$. (c) Optical conductivity $%
\text{Re}\left[ \protect\sigma _{xx}\right] $ as a function of $\hbar 
\protect\omega $. (d) Momentum $k_{x}$ as a function of $2E$ at $\left(
k_{y},k_{z}\right) =\left( 0,0\right) ,\left( 0,\protect\pi \right) ,\left( 
\protect\pi ,0\right) ,\left( \protect\pi ,\protect\pi \right) $. (e) DOS as
a function of $2E/t$.}
\label{Fig3DOpt}
\end{figure}

Finally, we study the tight-binding model on the cubic lattice, whose
Hamiltonian is given by\cite{Zhang,Liu}%
\begin{align}
H& =\left[ m_{0}-t\left( \cos k_{x}+\cos k_{y}+\cos k_{z}\right) \right]
\sigma _{z}  \notag \\
& +v\left( \sigma _{x}\sin k_{x}+\sigma _{y}\sin k_{y}+\sigma _{z}\sin
k_{z}\right) .
\end{align}%
It describes three-dimensional topological insulators\cite{Zhang,Liu} such
as Bi$_{2}$Se$_{3}$\ and Bi$_{2}$Te$_{3}$. The quantum metric is shown in
Fig.\ref{Fig3DOpt}(a1), (a2) and (b). The optical conductivity is shown in
Fig.\ref{Fig3DOpt}(c). The band-edge frequency of the optical conductivity
coincides with the band structure as in Fig.\ref{Fig3DOpt}(d). The optical
conductivity at the band-edge frequency is zero, which is consistent with
the Dirac model as shown in Fig.\ref{FigDiracOpt}(c2). The sharp peak in the
optical conductivity is due to the van-Hove singularity of the DOS as in Fig.%
\ref{Fig3DOpt}(e).

\section{Conclusion}

We have analytically determined the quantum metric and the optical
conductivity in the Dirac model with parabolic mass term in arbitrary
dimensions, and revealed that the parabolic dispersion of the Dirac mass
term quite affects the optical absorption. In addition, we have shown that
the optical absorption at the band-edge frequency exhibits a distinct
behavior depending on the dimension. We have studied two typical Chern
insulators, i.e., the QWZ model and the Haldane model. By comparing the
topological and trivial phases with the same gap, the optical absorption is
significantly different in these two phases in the QWZ model but not in the
Haldane model.

This work is supported by CREST, JST (Grants No. JPMJCR20T2) and
Grants-in-Aid for Scientific Research from MEXT KAKENHI (Grant No. 23H00171).

\appendix

\section{Quantum geometric tensor and quantum metric}

We review the relation between the optical conductivity and the quantum
metric\cite{Onishi}. The quantum distance is defined by\cite{Provost,Ma} 
\begin{align}
ds^{2}& =\sum_{nm}\left\vert \left\vert \psi _{n}\left( \mathbf{k}+\delta 
\mathbf{k}\right) -\psi _{m}\left( \mathbf{k}\right) \right\vert \right\vert
^{2}  \notag \\
& =\sum_{nm}\left\langle \psi _{n}\left( \mathbf{k}+\delta \mathbf{k}\right)
-\psi _{m}\left( \mathbf{k}\right) |\psi _{n}\left( \mathbf{k}+\delta 
\mathbf{k}\right) -\psi _{m}\left( \mathbf{k}\right) \right\rangle .
\end{align}%
Up to the second order, it is expanded as%
\begin{align}
ds^{2}& =\sum_{nm}\sum_{\mu \nu }\left\langle \partial _{k_{\mu }}\psi
_{n}\left( \mathbf{k}\right) dk_{\mu }\left\vert \partial _{k_{\nu }}\psi
_{m}\left( \mathbf{k}\right) dk_{\nu }\right\rangle \right. dk_{\mu }dk_{\nu
}  \notag \\
& =\sum_{nm}\sum_{\mu \nu }\mathcal{Q}_{\mu \nu }^{nm}\left( \mathbf{k}%
\right) dk_{\mu }dk_{\nu },
\end{align}%
where $Q_{\mu \nu }^{nm}\left( \mathbf{k}\right) $ is the quantum geometric
tensor, and given by\cite{Ma}%
\begin{equation}
\mathcal{Q}_{\mu \nu }^{nm}\left( \mathbf{k}\right) =\left\langle \partial
_{k_{\mu }}\psi _{n}\left( \mathbf{k}\right) \right\vert 1-P\left( \mathbf{k}%
\right) \left\vert \partial _{k_{\nu }}\psi _{m}\left( \mathbf{k}\right)
\right\rangle ,
\end{equation}%
with the projection operator%
\begin{equation}
P\left( \mathbf{k}\right) \equiv \sum_{n}\left\vert \psi _{n}\left( \mathbf{k%
}\right) \right\rangle \left\langle \psi _{n}\left( \mathbf{k}\right)
\right\vert .
\end{equation}%
The quantum geometric tensor is decomposed as%
\begin{equation}
\mathcal{Q}_{\mu \nu }^{nm}\left( \mathbf{k}\right) =g_{\mu \nu }^{nm}-\frac{%
i}{2}F_{\mu \nu }^{nm},
\end{equation}%
where%
\begin{equation}
g_{\mu \nu }^{nm}\equiv \frac{\mathcal{Q}_{\mu \nu }^{nm}+\mathcal{Q}_{\mu
\nu }^{nm\dagger }}{2}=\text{Re}\left[ \mathcal{Q}_{\mu \nu }^{nm}\right]
\end{equation}%
is the quantum metric, and 
\begin{equation}
F_{\mu \nu }^{nm}\equiv i\left( \mathcal{Q}_{\mu \nu }^{nm}-\mathcal{Q}_{\mu
\nu }^{nm\dagger }\right) =\text{Im}\left[ \mathcal{Q}_{\mu \nu }^{nm}\right]
\end{equation}%
is the non-Abelian Berry curvature.

\section{Quantum metric and optical conductivity}

The optical conductivity is calculated based on the Kubo formula as

\begin{align}
& \sigma _{\mu \nu }\left( \omega \right)  \notag \\
& =\frac{e^{2}}{\hbar }\int d\mathbf{k}\sum_{n,m}\left( f_{n}\left( \mathbf{k%
}\right) -f_{m}\left( \mathbf{k}\right) \right) \frac{\varepsilon
_{mn}\left( \mathbf{k}\right) A_{nm}^{\mu }\left( \mathbf{k}\right)
A_{nm}^{\nu }\left( \mathbf{k}\right) }{\varepsilon _{n}\left( \mathbf{k}%
\right) -\varepsilon _{m}\left( \mathbf{k}\right) +\hbar \omega +i\eta } 
\notag \\
& =\pi \omega e^{2}\int d\mathbf{k}\sum_{n,m}\left( f_{n}\left( \mathbf{k}%
\right) -f_{m}\left( \mathbf{k}\right) \right) A_{nm}^{\mu }\left( \mathbf{k}%
\right) A_{nm}^{\nu }\left( \mathbf{k}\right)  \notag \\
& \qquad \qquad \times \delta \left( \varepsilon _{n}\left( \mathbf{k}%
\right) -\varepsilon _{m}\left( \mathbf{k}\right) -\hbar \omega \right) ,
\end{align}%
where $A_{nm}^{\alpha }$ is the inter-band Berry connection defined as%
\begin{equation}
A_{nm}^{\mu }\left( \mathbf{k}\right) =i\left\langle \psi _{n}\left( \mathbf{%
k}\right) \right\vert \partial _{k_{\mu }}\left\vert \psi _{m}\left( \mathbf{%
k}\right) \right\rangle ,
\end{equation}%
while $f_{n}\left( \mathbf{k}\right) $ is the Fermi distribution function
and $\varepsilon _{n}\left( \mathbf{k}\right) $ is the band dispersion and $%
\eta $ is an infinitesimal real number.

Here, we have%
\begin{align}
& \sum_{nm}A_{nm}^{\mu }\left( \mathbf{k}\right) A_{mn}^{\nu }\left( \mathbf{%
k}\right)  \notag \\
& =-\sum_{nm}\left\langle \psi _{n}\left( \mathbf{k}\right) \right\vert
\partial _{k_{\mu }}\left\vert \psi _{m}\left( \mathbf{k}\right)
\right\rangle \left\langle \psi _{m}\left( \mathbf{k}\right) \right\vert
\partial _{k_{\nu }}\left\vert \psi _{n}\left( \mathbf{k}\right)
\right\rangle  \notag \\
& =\sum_{nm}\left\langle \partial _{k_{\mu }}\psi _{n}\left( \mathbf{k}%
\right) \left\vert \psi _{m}\left( \mathbf{k}\right) \right\rangle \right.
\left\langle \psi _{m}\left( \mathbf{k}\right) \left\vert \partial _{k_{\nu
}}\psi _{n}\left( \mathbf{k}\right) \right\rangle \right.  \notag \\
& =\sum_{nm}\left\langle \partial _{k_{\mu }}\psi _{n}\left( \mathbf{k}%
\right) \right\vert P_{m}\left\vert \partial _{k_{\nu }}\psi _{n}\left( 
\mathbf{k}\right) \right\rangle  \notag \\
& =\sum_{nm}\left\langle \partial _{k_{\mu }}\psi _{n}\left( \mathbf{k}%
\right) \right\vert \left( 1-P_{n}\right) -\left( 1-P_{n}-P_{m}\right)
\left\vert \partial _{k_{\nu }}\psi _{n}\left( \mathbf{k}\right)
\right\rangle  \notag \\
& =\mathcal{Q}_{\mu \nu }^{nn}\left( \mathbf{k}\right) ,
\end{align}%
where we have used the fact that the complete sum of the quantum metric over
the valence and conduction bands is zero, or%
\begin{equation}
\left\langle \partial _{k_{\mu }}\psi _{n}\left( \mathbf{k}\right)
\right\vert \left( 1-P_{n}-P_{m}\right) \left\vert \partial _{k_{\nu }}\psi
_{n}\left( \mathbf{k}\right) \right\rangle =0.
\end{equation}%
Then, taking the real part, we have%
\begin{equation}
\sum_{nm}\text{Re}\left[ A_{nm}^{\mu }\left( \mathbf{k}\right) A_{mn}^{\nu
}\left( \mathbf{k}\right) \right] =\sum_{n}g_{\mu \nu }^{nn}=g_{\mu \nu },
\end{equation}%
where we have defined%
\begin{equation}
g_{\mu \nu }=\sum_{n}g_{\mu \nu }^{nn}
\end{equation}%
and the real part of the optical conductivity is calculated as%
\begin{align}
& \text{Re}[\sigma _{xx}\left( \omega \right) ]  \notag \\
& =\pi \omega e^{2}\int d\mathbf{k}\sum_{n,m}\left( f_{n}\left( \mathbf{k}%
\right) -f_{m}\left( \mathbf{k}\right) \right)  \notag \\
& \qquad \qquad \times \text{Re}\left[ A_{nm}^{\alpha }\left( \mathbf{k}%
\right) A_{nm}^{\beta }\left( \mathbf{k}\right) \right] \delta \left(
\varepsilon _{n}\left( \mathbf{k}\right) -\varepsilon _{m}\left( \mathbf{k}%
\right) -\hbar \omega \right)  \notag \\
& =\pi \omega e^{2}\int d\mathbf{k}\sum_{n,m}\left( f_{n}\left( \mathbf{k}%
\right) -f_{m}\left( \mathbf{k}\right) \right)  \notag \\
& \qquad \qquad \times \text{Tr}\left[ g_{xx}\right] \delta \left(
\varepsilon _{n}\left( \mathbf{k}\right) -\varepsilon _{m}\left( \mathbf{k}%
\right) -\hbar \omega \right) .
\end{align}%
This is Eq.(\ref{Conduc}) in the main text.

\section{Detailed derivation of quantum metric}

We obtain 
\begin{align}
& \sum_{j}\left( \partial _{k_{x}}n_{j}\right) ^{2}  \notag \\
& =\left( \frac{2uk_{x}}{E}-\frac{m+uk^{2}}{E^{3}}k_{x}\mathcal{E}\right)
^{2}+\left( \frac{v}{E}-\frac{vk_{x}^{2}}{E^{3}}\mathcal{E}\right) ^{2} 
\notag \\
& +\sum_{j=2}^{N}\left( -\frac{vk_{j}k_{x}}{E^{3}}\mathcal{E}\right) ^{2},
\end{align}%
where we have introduced%
\begin{equation}
\mathcal{E=}v^{2}+2mu+2u^{2}k^{2}.
\end{equation}%
We further obtain%
\begin{eqnarray}
&&\sum_{j}\left( \partial _{k_{x}}n_{j}\right) ^{2}  \notag \\
&=&\frac{1}{E^{6}}[\left( 2uk_{x}E^{2}\right) ^{2}+\left( m+uk^{2}\right)
^{2}k_{x}^{2}\mathcal{E}^{2}  \notag \\
&&-4uk_{x}E^{2}\left( m+uk^{2}\right) k_{x}\mathcal{E}+\left( vE^{2}\right)
^{2}  \notag \\
&&+\left( vk_{x}^{2}\mathcal{E}\right) ^{2}-2vE^{2}vk_{x}^{2}\mathcal{E+}%
\sum_{j=2}^{N}\left( vk_{j}k_{x}\mathcal{E}\right) ^{2}  \notag \\
&=&\frac{1}{E^{6}}\left( vk_{x}\mathcal{E}\right)
^{2}\sum_{j=1}^{N}k_{j}^{2}+\frac{\left( 2uk_{x}\right) ^{2}+v^{2}}{E^{2}} 
\notag \\
&&-\frac{2\mathcal{E}^{2}k_{x}^{2}}{E^{4}}+\frac{\left( m+uk^{2}\right)
^{2}k_{x}^{2}\mathcal{E}^{2}}{E^{6}}  \notag \\
&=&\frac{k_{x}^{2}\mathcal{E}^{2}}{E^{6}}\left( v^{2}k^{2}+\left(
m+uk^{2}\right) ^{2}\right) +\frac{\left( 2uk_{x}\right) ^{2}+v^{2}}{E^{2}}-%
\frac{2\mathcal{E}^{2}k_{x}^{2}}{E^{4}}  \notag \\
&=&\frac{k_{x}^{2}\mathcal{E}^{2}}{E^{6}}E^{2}+\frac{\left( 2uk_{x}\right)
^{2}+v^{2}}{E^{2}}-\frac{2\mathcal{E}^{2}k_{x}^{2}}{E^{4}}  \notag \\
&=&-\frac{k_{x}^{2}\mathcal{E}^{2}}{E^{4}}+\frac{\left( 2uk_{x}\right)
^{2}+v^{2}}{E^{2}}  \notag \\
&=&\frac{v^{2}}{E^{2}}\left( 1-k_{x}^{2}\frac{4mu+v^{2}}{E^{2}}\right) .
\end{eqnarray}%
Hence, the quantum metric is given by 
\begin{equation}
g_{xx}\left( \mathbf{k}\right) =\frac{2^{N-3}v^{2}}{E^{2}}\left( 1-k_{x}^{2}%
\frac{4mu+v^{2}}{E^{2}}\right) .
\end{equation}%
This is Eq.(\ref{gxxk}) in the main text.

\section{($N$-1)-sphere}

We summarize the ($N$-1)-sphere coordinate. The momenta are parametrized as%
\begin{align}
k_{1}& =k\cos \theta _{1},  \notag \\
k_{2}& =k\sin \theta _{1}\cos \theta _{2}  \notag \\
k_{3}& =k\sin \theta _{1}\sin \theta _{2}\cos \theta _{3}  \notag \\
k_{4}& =k\sin \theta _{1}\sin \theta _{2}\sin \theta _{3}\cos \theta _{4} 
\notag \\
& \cdots  \notag \\
k_{N}& =k\sin \theta _{1}\sin \theta _{2}\sin \theta _{3}\cdots \sin \theta
_{N-1},
\end{align}%
where $0\leq k\leq \infty $, $0\leq \theta _{j}\leq \pi $ for $1\leq j\leq
N-2$ and $0\leq \theta _{N-1}\leq 2\pi $. The Jacobian is given by%
\begin{equation}
J=k^{N-1}\sin ^{N-2}\theta _{1}\sin _{2}^{N-3}\theta \cdots \sin \theta
_{N-2}dkd\theta _{1}d\theta _{2}\cdots d\theta _{N-1}.
\end{equation}%
The area of the ($N$-1)-sphere is given by 
\begin{equation}
\int_{0}^{\pi }d\theta _{1}\int_{0}^{\pi }d\theta _{2}\cdots \int_{0}^{\pi
}d\theta _{N-2}\int_{0}^{2\pi }d\theta _{N-1}J=\frac{N\pi ^{N/2}}{\Gamma
\left( \frac{N}{2}+1\right) }.
\end{equation}%
It is $2$ for $N=1$, $2\pi $ for $N=2$ and $4\pi $ for $N=3$. On the other
hand%
\begin{align}
& \int_{0}^{\pi }d\theta _{1}\int_{0}^{\pi }d\theta _{2}\cdots \int_{0}^{\pi
}d\theta _{N-2}\int_{0}^{2\pi }d\theta _{N-1}J\frac{k_{x}^{2}}{k^{2}}  \notag
\\
& =\int_{0}^{\pi }d\theta _{1}\int_{0}^{\pi }d\theta _{2}\cdots
\int_{0}^{\pi }d\theta _{N-2}\int_{0}^{2\pi }d\theta _{N-1}J\cos ^{2}\theta
_{1}  \notag \\
& =\frac{N\pi ^{N/2}\Gamma \left( \frac{N}{2}\right) }{2\Gamma \left( \frac{N%
}{2}+1\right) ^{2}}=\frac{\pi ^{N/2}}{\Gamma \left( \frac{N}{2}+1\right) }.
\end{align}%
It is $2$ for $N=1$, $\pi $ for $N=2$ and $4\pi /3$ for $N=3$. They are used
in the integration of $g\left( \mathbf{k}\right) $ to derive $g\left(
k\right) $ in Eq.(\ref{gxx}).

\end{document}